\documentclass[prl,twocolumn,showpacs,superscriptaddress]{revtex4}
\usepackage{graphicx}

\begin{document}  

\title {Time-resolved dynamics of electron wave packets in chaotic and
  regular quantum billiards with leads}

\author{I. V. Zozoulenko}\email{igozo@itn.liu.se} 
\affiliation{
Department of Science and Technology (ITN), Link\"{o}ping University, 
S--601\,74 
Norrk\"oping, Sweden}
\affiliation{Department of Physics (IFM),
Link\"{o}ping University, S--581\,83 Link\"{o}ping, Sweden}

\author{T. Blomquist}\email{torbl@ifm.liu.se} 
\affiliation{Department of Physics (IFM),
Link\"{o}ping University, S--581\,83 Link\"{o}ping, Sweden}
\date{\today}

\begin{abstract}
We perform numerical studies of the wave packet propagation through open quantum 
billiards whose
classical counterparts exhibit regular and chaotic dynamics. We show
that for $t\lesssim\tau_H$ ($\tau_H$ being the Heisenberg time), the features in 
the transmitted and
reflected currents are directly related to specific classical
trajectories connecting the billiard leads. In contrast, the long-time
asymptotics of the wave packet dynamics is qualitatively different for
classical and quantum billiards. In particularly, the decay of the quantum 
system  
obeys a power law that depends
on the number of decay channels, and is not sensitive to the  nature 
of classical dynamics (chaotic or regular).  

\end{abstract} 

\maketitle

Low-dimensional nanometer-scaled semicon\-duc\-tor struc\-tures, quantum dots
(sometimes called the quantum electron billiards) represent artificial man-made
systems which are 
well suited to study different aspects of
quantum-mechanical scattering \cite{Marcus_Andy}. 
A majority of studies of electron transport in such systems have
been mainly focused on the stationary electron dynamics. In recent years,
however, interest in temporal aspect of quantum scattering  has been 
renewed \cite{Heller1}. This includes e.g., studies of
the time delay distributions \cite{Fyodorov} and correlation decay in quantum 
billiards and 
related systems \cite{Ben}. 
Furthermore,
 many core starting points
in the description of the stationary scattering rely heavily on the properties
of the system in the time domain.
In particular, the semiclassical approach exploits the difference between the 
classical escape rate from the cavities with chaotic and regular (or mixed) 
dynamics (exponentially fast $e^{-\gamma t}$ for the former vs. power-law 
$t^{-\xi}$ for the later \cite{Bauer}). This difference in the classical 
dynamics 
translates into the the difference in observed transport properties (statistics 
of the fluctuations 
\cite{Smilansky,Jalabert,Ker}, a shape 
of the weak localization \cite{Baranger}, etc.).

On the other hand, the quantum mechanical (QM) approaches predict qualitatively 
different, universal power-law escape rate from the cavity 
\cite{Muller}, 
\begin{eqnarray}\label{power}
d\, P(t)/\,d\,t\ \sim\ t^{-\beta \frac{M}{2}-1},
\end{eqnarray} where $P(t)$ is the survival probability, $M$ is the number of 
decay channels and $\beta =1\ (2)$ for the system with (without) time-reversal 
invariance \footnote{Note that the specific power of the decay law (\ref{power}) 
depends on initial population of the states \protect \cite{Muller} as
well as on the strength of the coupling}. The 
non-exponential decay of a 
quantum system with chaotic classical dynamics has been indirectly demonstrated 
 \cite{Alt} in a microwave stadium billiard.

The QM power-law delay time for the chaotic cavity is expected to deviate from the 
semi-classical (SC) decay at times of the order of the Heisenberg time $ 
\tau_H=\hbar/\Delta, \ \Delta $ being the mean level spacing of the cavity 
\cite{Lewenkopf}. At 
the same time, the difference in the classical decay of the chaotic and 
regular/mixed cavities often becomes discernible only after many bounces at the 
times 
which often exceed $\tau_H$ \cite{TB2}. Nevertheless, the SC predictions are 
widely used 
in 
experiment to distinguish between the chaotic and regular/mixed dynamics in 
quantum billiards \cite{Marcus_Andy}. Is it thus possible to reconcile the SC 
and QM approaches, 
or should 
some of the SC predictions be used with certain caution or even be revised? 
Does the long-time decay asymptotics of the \textit{quantum} systems depend on 
the underlying \textit{classical} dynamics (chaotic or regular)?
Motivated by these questions we, in this paper, perform direct quantum
mechanical calculations of the passage of electron wave packets through 
two-dimensional electron billiards. 

To the best of our knowledge, all of the studies of wave packets dynamics in 
{\it open} systems presented so far, have been mostly restricted to (a) quantum 
limit where a characteristic size of the system $L$ was of the order of the 
average wavelength of the wave packet $\langle \lambda \rangle$ and (b) to an 
initial stage of the wave packet evolution
$t\lesssim 1$ (where $t$ is in units of the traversal time). 
The time-dependent solution of the Schr\"odinger 
equation was typically obtained on the basis of direct schemes  approximating 
the exponential time propagator \cite{time}. With such methods the task of 
tracing the 
long-time evolution of a wave packet in a realistic quantum dot would be 
forbiddingly expensive in terms of both computing power and memory. In the 
present paper we thus implement a spectral method based on the Green function 
technique  \cite{Hauge}, which allows us (a) to reach a semi-classical regime 
$\langle \lambda \rangle \gg L$ and (b) to approach a long-time asymptotics 
$t\gg 1$ corresponding to $10^4 - 10^5$ bounces of a classical particle in a 
billiard. We found that during the initial phase $t\lesssim \tau_H$ (which in 
our case corresponds to  $\sim 10 - 20$ classical bounces in a billiard), the 
QM decay closely follows the classical one, such that all the features in the 
QM 
current leaking out of the billiard can be explained in terms of 
geometry-specific classical trajectories between the leads. When $t \gtrsim 
\tau_H$, the QM dynamics starts to deviate from the classical one, with the 
decay rate obeying a power law that depends
on the number of decay channels only,  irrespective of the nature 
of classical dynamics (chaotic or regular).  
We thus conclude that quantum mechanics smears out the difference between 
classically chaotic and regular motion.

\begin{figure} 
\includegraphics[width=0.3\textwidth]{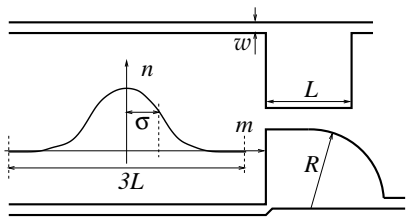}
\caption{\label{fig1}
A square and a quarter-stadium shaped billiard connected to semi-infinite leads; 
$L=R, L/w=8$. The half-width of the wave packet $\sigma=0.4L$; at $t=0$ the 
wave packet is distinct from zero in the interval of $3L$. The average 
wavelength  of the wave packet $\langle \lambda\rangle=2\pi/\langle 
k\rangle=0.8w$.
 } 
\end{figure}

We have studied the temporal evolution of wave packets in square, Sinai, and 
stadium billiards of various shapes. All of them exhibit similar features and 
we 
thus present here the results for two representative geometries, a square 
(which 
is classically regular) and a quarter-stadium (which is classically chaotic), 
see 
Fig. \ref{fig1}. The billiards are connected to two semi-infinite leads that 
can support one or more propagating modes. Magnetic field is restricted to 
zero. We assume a hard wall 
confinement both in the leads and in the interior of billiards. 
Dynamics of the wave packet is governed by the time-dependent Schr\"odinger 
equation
\begin{eqnarray}\label{Schrodinger}
\left(i\hbar\frac{\partial}{\partial t} -H\right)|\psi(t)\rangle =0,
\end{eqnarray}
where $H$ is the Hamiltonian operator and $|\psi(t)\rangle $ is the wave 
function. To study the time evolution of the initial state we follow St\o
vneng and Hauge \cite{Hauge} and perform the Laplace transform of Eq. 
(\ref{Schrodinger}) followed by the integration by parts. Changing variables in 
the Mellin inversion integral we obtain
\begin{eqnarray}\label{psi}
|\psi(t)\rangle 
=\frac{i}{2\pi}\int^{\infty+i0}_{-\infty+i0}dz\,G(z)|\psi(0)\rangle 
e^{-izt/\hbar},
\end{eqnarray}
where  we have introduced the Green function operator 
$G(z)=(z-H)^{-1}$ and taken into account that all the poles of the Green 
function lie in the lower $z$-plane.
With the help of Eq. (\ref{psi}), the calculation of the temporal evolution of 
the initial state is effectively reduced to the computation of the Green 
function of the Hamiltonian operator $H$ in the energy domain. To compute the 
Green function we discretize the system under consideration, introduce a 
standard tight-binding Hamiltonian and make use of the modified 
recursive Green-function technique described in details in  \cite{Z}.

Let us consider a minimum-uncertainty wave packet of the average energy $E$ 
which 
enters a billiard from the left lead in one of the transverse 
modes $\alpha$. We thus write the initial state in the left lead at $t=0$ in 
the 
form
\begin{eqnarray}
\label{packet}
|\psi_\alpha(0)\rangle&=&\sum_{mn}
\phi_m^\alpha
f^{n}_{\alpha}|mn\rangle,\\
\label{phi}
\phi_m^\alpha
&=&\frac{1}{(2\pi)^{1/4}\sqrt{\sigma}}
e^
{-\frac{(m-m_0)^2}{4\sigma^2}+ik_\parallel^\alpha m}
\end{eqnarray}
where $w$ is the width of the leads (measured in units of a lattice constant 
$a$),  
$f^{n}_{\alpha}=\sqrt{2/w}\sin(\pi\alpha n/w)$ is the eigenfunction of the 
transverse motion; 
$\langle k \rangle=2\pi/\langle \lambda \rangle=\sqrt{E/u}$ is the average wave 
vector (in units of $a^{-1}$), where $ u=\hbar^2/2m^*a^2$ and  $m^*$ being the 
effective mass; $\langle k 
\rangle^2={k_\parallel^\alpha}^2+{k_\perp^\alpha}^2$, 
$k_\perp^\alpha=\pi\alpha/w$, with $k_\parallel^\alpha$ and 
$k_\perp^\alpha$ being the longitudinal and transverse wave vectors 
respectively. The matrix 
element $\langle
m,n|\psi\rangle$ defines the probability amplitude to find
the electron on the site ($m,n$).  After the wave packet enters the billiard, it 
will leak out 
through both of the leads in all the available modes $\beta$. The wave function 
in e.g. the 
right lead can then be written in the form
\begin{eqnarray}
\label{lead}
|\psi_\alpha(t)\rangle=\sum_{m\beta}c_{\beta\alpha}^m(t)
|m\beta\rangle,
\end{eqnarray}
where $|m,\alpha\rangle=\sum_nf_\alpha^m|mn\rangle$; $c_{\beta\alpha}^m$ gives a 
probability to find a particle on the slice 
$m$ in the transverse mode $\beta$, provided the initial state enters the 
billiard in the mode $\alpha$. Discretizing a standard expression for the 
quantum-mechanical current, 
$j(x,y)=i\hbar/2m^*\left(\psi\nabla\psi^*-\psi^*\nabla\psi\right)$ we, using 
Eq. 
(\ref{lead}), obtain the following expression for the total current $J=\int\, 
dy\, j(x,y)$ through the slice $m$ in the leads expressed via coefficients 
$c_{\beta\alpha}^m(t)$ 
\begin{eqnarray}
\label{current}
J=\frac{i\hbar}{m^*a}\sum_\beta\left[c_{\beta\alpha}^m\left({c_{\beta\alpha}^{m
+1}}^*-{c_{\beta\alpha}^{m-1}}^*\right)-\text{c.c.}\right].
\end{eqnarray}
Note that the quantum-mechanical current is related to the survival probability 
in the billiard Eq. (\ref{power}) by the obvious relation
\begin{eqnarray}
\label{P(t)}
d\, P(t)/\,d\,t\, = - J_l(t)-J_r(t),
\end{eqnarray}
where $J_l(t)$ and $J_r(t)$ stand for the currents flowing into the left and 
right leads respectively. Note that the function 
$J_l(t)+J_r(t)$ has a meaning of the distribution of the time delays in the 
billiard \cite{Fyodorov}
We calculate coefficients ${c_{\beta\alpha}^{m+1}}$ in Eq. (\ref{current}) by 
computing a matrix 
element $\langle m\beta|\psi\rangle$ using Eqs. (\ref{psi})-(\ref{lead})
\begin{eqnarray}
\label{c}
c_{\beta\alpha}^m(t)=
\frac{i}{2\pi}\int^{\infty+i0}_{-\infty+i0}dz\,\sum_{m'}G_{\beta\alpha}^{mm'}(z
)
\phi_m^\alpha e^{-izt/\hbar},
\end{eqnarray}
where $G_{\beta\alpha}^{mm'}(z)$ stands for the matrix element  $\langle 
m\beta|G(z)|m'\alpha\rangle$ of the Green 
function of 
the whole system (billiard and semi-infinite leads).

A special care has been taken to ensure the reliability of the results of the  
numerical simulations. 
This, in particularly, includes the thorough control of the conservation of 
the total current. We have also calculated the temporal evolution of the wave 
packet in the infinite lead of the width $w$ and found an excellent agreement 
with the analytical results for a 1D lattice  \cite{Hauge}. 
Finally, the developed method reproduces correctly the conductance quantization 
of the quantum point contacts.

\begin{figure} 
\includegraphics[width=0.4\textwidth]{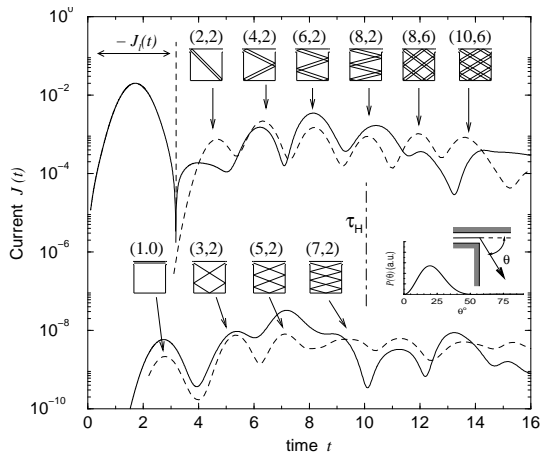}
\caption{\label{fig2}
Quantum mechanical reflected and transmitted current through a square billiard 
(upper and lower solid curves respectively). Dashed lines indicate 
corresponding {\it classical} currents. The wave packet enters the billiard in 
the second mode $\alpha=2$; $\langle k\rangle=2.5w/\pi$. The insets show 
classical reflected and transmitted trajectories; the
numbers in the parenthesis are the so-called winding numbers indicating how many 
times an electron traverses the billiard in the longitudinal and transverse 
directions. The inset to the right show an angular distribution of injected 
electrons $P(\theta)$ calculated in the Fraunhofer approximation for 
$\alpha=2$. The Heisenberg time $\tau_H$ is indicated by a dot-dashed line.
The curves are 
shifted for clarity.} 
\end{figure}

Let us first concentrate on the initial phase of the wave packet dynamics. 
Figure \ref{fig2} shows the quantum-mechanical current for the square billiard 
in the time interval $t\lesssim 15$ (here and hereafter we measure time in 
units of the traversal time $t_{tr}=L/\langle v \rangle$, $\langle v 
\rangle=L/(\hbar \langle k \rangle/m^*)$; the current 
 is measured in units of $\hbar/2m^*a$).  Our choice of the parameter of 
the wave packet $\sigma$ and $\langle k\rangle$ ensures that the spreading of 
the wave packet becomes noticeable only after relatively long time $t\gtrsim 
50$. 
The initial period of time $0<t\lesssim 3$, when the current through the left 
lead 
$J_l(t)$ 
is negative,  corresponds to a buildup phase when 
the 
wave packet enters the billiard. 
Having entered the billiard, the wave packet starts to 
leak out through the leads and the calculated QM currents $J_l, J_r$ show a 
series of pronounced peaks. To outline the origin of these peaks we 
calculate 
the leakage current of a  corresponding \textit{classical} wave packet in the 
same billiard. In the classical calculations we take into account the 
diffractive effects in the leads in the framework of the standard Fraunhofer 
diffraction approximation by injecting the electrons with the corresponding 
angular distribution $P(\theta)$ (inset in Fig. \ref{fig2} shows a calculated 
angular 
probability distribution for the lead geometry under consideration). 
A very good correspondence between the QM and the classical results allows us 
to ascribe each peak 
in the \textit{quantum-mechanical} transmitted/reflected currents  to a 
specific  \textit{classical} trajectory  connecting the billiard leads, see 
inset in Fig. 
\ref{fig2}.

\begin{figure} 
\includegraphics[width=0.4\textwidth]{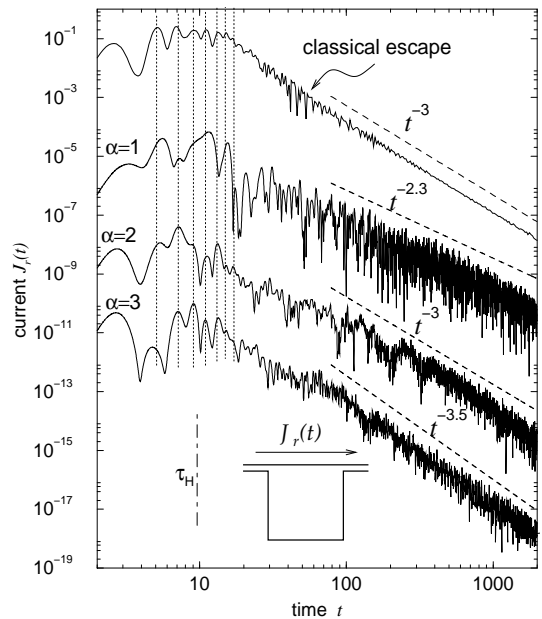}
\caption{\label{fig3}
Quantum mechanical transmitted current through a square billiard for different 
incoming modes of the wave packet $\alpha=1, 2, 3$ with  $\langle k\rangle \pi 
/w=1.5, 2.5, 3.5$ respectively. The upper curve indicates corresponding {\it 
classical} current. Vertical dotted lines are guides for an eye indicating the 
same positions of the peaks in the current. Dashed lines give the asymptotic 
power-law decay obtained by the best fit in the interval $80<t<2000$. The 
Heisenberg time $\tau_H$ is indicated by a dot-dashed line. The curves are 
shifted for clarity. } 
\end{figure}

A relative height of each peak depends on the density of the corresponding 
trajectories, and the angular distribution for a given incoming mode $\alpha$. 
The 
effect of the later is clearly seen in Fig. \ref{fig3}, where the QM current in 
a square billiard was shown for three different incoming modes of the wave 
packet $\alpha=1, 2, 3$. At the initial stage of the current decay, $t\lesssim 
20$, the positions of peaks are the same for all incoming modes, whereas their 
absolute values are different. This is explained by the fact that the angular 
distribution $P(\theta)$ is different for different $\alpha$, with its maximum 
being shifted to larger $\theta$ for higher modes $\alpha$.
We conclude this discussion by noticing that  all quantum billiards
studied here exhibit similar characteristic peaks in the current at $t<\tau_H$ 
that can be
explained in terms of corresponding classical trajectories.

\begin{figure} 
\includegraphics[width=0.4\textwidth]{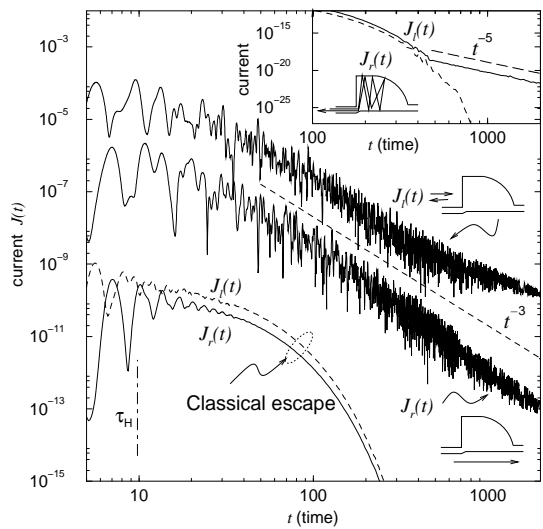}
\caption{\label{fig4}
Quantum mechanical reflected and transmitted current through a
stadium-shaped billiard. Lower curves show 
corresponding {\it classical} currents. The wave packet enters the billiard in 
the second mode $\alpha=2$; $\langle k\rangle=2.5w/\pi$. The inset show 
a long-time assymptotics of classical escape along with the example of
a bouncing-ball trajectory.  Dashed lines give the asymptotic 
power-law decay obtained by the best fit in the interval $50<t<2000$. The 
Heisenberg time $\tau_H$ is indicated by a dot-dashed line.
Time is measured in the units of the traversal time of the equivalent
square of the same area. The curves are 
shifted for clarity. } 
\end{figure}

Let us now focus on a long-time asymptotics of the wave packet dynamics. We 
start with a square billiard, see Fig. \ref{fig3}. Its classical escape rate is 
independent of the number of modes in the leads and is well approximated by a 
power law $\sim t^{-3}$.
On the contrary, the calculated quantum-mechanical decay, does depend on the 
number of modes in the leads  and  follows the power law decay with 
the exponents $\xi=2.3,3,3.5$  for modes $\alpha=1,2,3$ correspondingly. These 
values are somehow different from those expected from Eq. (\ref{power}), 
$\xi=2,3,4$. (In a billiard with two leads a number of decay 
channels is given by $M=2\alpha$).

One of the reasons for the above discrepancy may be related to the
fact that in a billiard system the details of the coupling between the
leads can be important for the selection of particular states that
mediate transport through the system \cite{eigen} (on the contrary, Eq. 
(\ref{power}) corresponds to 
the case when all resonant  states are excited with the same 
probability at $t=0$ ).
Note that eq.~(\ref{power}) corresponds to the
weak (tunneling) coupling between the leads and the dots in contrast
to the regime of the open dot considered here.  
It is also worth to mention that  
Eq. (\ref{power}) is based on the random matrix theory and similar stochastic 
approaches \cite{Lewenkopf,Muller,Fyodorov}. The predictions of these theories 
tend to be rather general in 
nature and they usually fail to account for specific features of the geometry 
under consideration (such as  details of the lead position, existence of 
periodic orbits, etc.).

Let us now discuss wave-packet evolution in  a stadium-shaped
billiard. Classically, this billiard exhibits chaotic
dynamics, and its classical escape rate shows fast exponential
decay, see Fig. \ref{fig4}. The quantum mechanical decay of such a
system is however, qualitatively different. It obeys a
power law  similar to the one observed for the square. The
difference between the classical and QM decay asymptotics becomes
clearly discernible at the time scale  corresponding to $\sim 50-100$
classical bounces in a billiard. With this respect it is
important to stress that the difference in the power-law and the exponential 
escape for \textit{classical} regular and chaotic systems becomes discernible at 
a comparable time interval\cite{TB2}. Note that the billiard at hand is designed 
in
such a way that the classical escape through the left lead changes
its asymptotics from the exponential one to a slower power-law decay at $t\sim
500$. This behavior is caused by the bouncing-ball
orbits \cite{Alt2} which are accessible via the left lead only (see inset in
Fig. \ref{fig4}). It is interesting to note that the corresponding QM current
through the left lead also starts to show slower decay at $t\sim
500$ in comparison to the right lead. We therefore
speculate that, even though the long-time asymptotics of the QM and
classical decays are qualitatively different, the QM decay still
reflects some features of the underlying classical dynamics.

The qualitative difference between the QM and classical escape represents
one of the main findings of the present work. We also find that the
asymptotic decay of a quantum system  obeys a power law that depends
on the number of decay channels only, and is not sensitive to the  nature 
of classical dynamics (chaotic or regular). 
This makes us conclude that quantum mechanics smears out the difference 
between 
classical chaotic and regular motion. With this respect it is
important to stress that the difference in the \textit{classical}
decay rate in chaotic, regular or mixed system is often used in
various semiclassical approaches to describe observed transport
properties of the quantum systems (statistics of the fluctuations, the shape 
of weak localization, fractal
conductance fluctuations,  etc.) 
\cite{Marcus_Andy,Smilansky,Jalabert,Baranger,Ker}. We demonstrate however, that 
the crossover to the QM power law decay may occur at 
the same time scale when the difference between \textit{classical} regular and 
chaotic systems becomes discernible.  
Our findings  thus strongly indicate
that some of the SC predictions should be used with certain caution or
even be revised. Finally, the results reported in the present paper can
be tested experimentally in the variety of systems including
semiconductor quantum dots \cite{Marcus_Andy}, microwave
cavities \cite{Stoeckmann}, acoustical \cite{acoustics} and optical
billiards \cite{optics}.

\acknowledgments Financial support of NFR and VR (I.V.Z) and NGSSC
(T.B.) is greatly acknowledged.

\end{document}